\documentclass[
 reprint,
 amsmath,amssymb,
 aps,
 floatfix,
]{revtex4-2}
\usepackage{enumerate}
\usepackage{amsmath}
\usepackage{amsfonts}
\usepackage{amssymb}
\usepackage[utf8]{inputenc}
\usepackage[T1]{fontenc}
\usepackage{mathtools}
\usepackage{wasysym}
\usepackage{accents}
\usepackage{graphicx}
\usepackage[english]{babel}
\usepackage[svgnames]{xcolor}
\usepackage[colorlinks=true,citecolor=blue]{hyperref}
\usepackage{url}
\usepackage{amsmath}
\usepackage{float}
\usepackage{mhchem}
\usepackage{bbold}
\usepackage{subcaption}
\usepackage{dcolumn}
\usepackage{bm}

\begin{document}

\preprint{MDR-WD/2026}

\title{Modified electron dispersion relations in degenerate white dwarfs}

\author{Edson Otoniel}
\affiliation{Instituto de Formação de Educadores , Universidade Federal do Cariri, R. Olegário Emidio de Araujo, s/n - Centro, Brejo Santo - CE, 63260-000, Brazil}

\author{Iarley P. Lobo}
\affiliation{Physics Department,  Federal University of Paraíba, Caixa Postal 5008, 58051--970, João Pessoa, Paraíba, Brazil.}
\affiliation{Physics Department,  Federal University of Campina Grande, Caixa Postal 10071, 58429-900 Campina Grande, Paraíba, Brazil.}

\date{\today}

\begin{abstract}
We investigate how modified electron dispersion relations affect the structure of cold white dwarfs (WDs). The deformation is introduced only in the degenerate electron equation of state, through the energy of a single particle and the group velocity, while the stellar mass density remains dominated by ions through the relation $\rho\simeq \mu_e m_u n_e$ imposed by charge neutrality. The resulting equations of state are coupled to the standard TOV equations, with no modification of the gravitational field equations. For a baseline composed of carbon and oxygen with $\mu_e=2$, the undeformed limit recovers a maximum mass in the Chandrasekhar scale, validating the normalization of the calculation before the modified cases are considered. The first model of modified dispersion produces only a modest stiffening over the parameter range studied, whereas the logarithmic model depends strongly on the sign of the deformation parameter: negative values increase the pressure and the maximum mass, while positive values soften the sequence. These results show that WDs can isolate the impact of modified electron kinematics on compact star structure, but the logarithmic branch in particular requires further restrictions from its physical domain, stability conditions, and observational constraints on mass and radius.
The calculation implements the modified dispersion relation in the electron matter sector while retaining the Einstein field equations and the standard general relativistic description of hydrostatic equilibrium.
\end{abstract}

\maketitle

\section{Introduction}

Compact stars connect microscopic particle physics with macroscopic gravitational equilibrium. In neutron stars this connection is obscured by the uncertain dense nuclear equation of state, whereas in white dwarfs (WDs) the dominant pressure is produced by a cold degenerate electron gas and the rest mass density is carried mainly by ions \cite{Glendenning2000}. This separation makes WDs a controlled environment for asking how changes in the electron relation between energy and momentum propagate into an equation of state and then into a relation between mass and radius. The undeformed reference point is the Chandrasekhar model, in which relativistic electron degeneracy pressure fixes the characteristic maximum mass scale for a given mean molecular weight per electron \cite{Chandrasekhar:1931ih}. Modern relativistic treatments refine this baseline by including general relativistic structure, finite nuclear size, Coulomb effects, rotation, and modified gravity corrections \cite{rotondo_relativistic_2011,boshkayev_general_2013,otoniel_white_2026}.

Different approaches to quantum gravity suggest that quantum-spacetime effects modify the kinematics of particles through corrections that are suppressed by the Planck scale \cite{Amelino-Camelia:2008aez}. These effects are commonly described by modified dispersion relations, whose precise form depends on the underlying quantum-gravity framework \cite{Jacobson:2005bg,Liberati:2013xla}.Modified dispersion relations provide a phenomenological way to parameterize departures from the special relativistic relation between energy and momentum. They appear in approaches to deformed or doubly special relativity, Lorentz symmetry violation, and quantum spacetime phenomenology, where an observer independent energy scale may supplement the invariant speed of light \cite{Amelino-Camelia:2000stu,Magueijo:2001cr,Colladay:1998fq}. Broader phenomenological studies connect these ideas to quantum spacetime tests and high energy constraints \cite{Amelino-Camelia:2008aez,Addazi:2021xuf,AlvesBatista:2023wqm}. Such deformations have been studied in gamma ray propagation, reaction thresholds, black hole thermodynamics, and cosmology \cite{Amelino-Camelia:1997ieq,Jacobson:2002hd,Ali:2014xqa}. Related applications to degenerate Fermi systems motivate their use in compact objects \cite{Santos:2015sva,Amelino-Camelia:2009iag}. For compact objects, the relevant question is not only whether the single particle energy $E(k)$ changes, but whether the associated group velocity $dE/dk$ changes the pressure integral enough to produce measurable effects on stellar structure.

In deformed relativistic kinematics, the functions that modify the mass shell are defined in momentum space. Their presence does not, by itself, select a unique energy dependent metric for the macroscopic spacetime. Momentum dependent spacetime metrics, including the construction known as rainbow gravity, require an additional prescription for extending the deformed kinematics to gravity \cite{Magueijo:2002xx,Lobo:2017rainbows,Pfeifer:2022drk}. In this work we adopt a matter sector implementation: the electron dispersion relation is modified in the local frame of the stellar fluid, whereas the spacetime dynamics is governed by general relativity.

The immediate starting point for the present work is the modified free Fermi gas model of Santos \textit{et al.}, who coupled deformed equations of state to the TOV equations for a simplified neutron star model composed of noninteracting neutrons \cite{Santos:2021avc}. Their calculation established a useful workflow: choose a modified dispersion relation, compute the corresponding Fermi gas energy density and pressure at zero temperature, and solve the relativistic equations of stellar structure. Previous studies have also discussed modified dispersion relations in the context of the Chandrasekhar mass and WD limits \cite{Gregg:2008jb,Amelino-Camelia:2009iag}. Massive WDs have been explored through magnetic support, modified gravity, and rotation \cite{otoniel_fermionic_2015,otoniel_white_2017,carvalho_stellar_2017,otoniel_strongly_2019}, while nuclear stability limits provide additional reference points for any mechanism that raises the WD mass scale \cite{chamel_stability_2013,chatterjee_maximum_2017}. However, extending the free Fermi gas construction to WDs requires an additional physical step. One cannot simply replace the neutron mass by the electron mass throughout the stellar model, because electrons supply the degeneracy pressure while ions dominate the mass density.

The central physical simplification of the present work is to isolate the electron sector before introducing the additional ingredients of a realistic WD model. In this sense, the deformation is not treated as a change in the gravitational field equations, but as a possible modification of the microscopic kinematics that produces the degenerate pressure. The ionic component still sets the dominant rest mass density, while the electrons determine how efficiently the matter resists compression. This choice provides a controlled first step: it separates the effect of modified electron kinematics from other mechanisms that can also alter massive WDs, such as finite temperature, crystallization, Coulomb corrections, magnetic fields, rotation, and inverse beta decay. These effects are essential in realistic electron ion plasma equations of state and in the stability of massive WD configurations \cite{chabrier_equation_1998,potekhin_equation_2000,potekhin_equation_2013,chamel_electron_2015}. Rotation and magnetic deformation can further modify the structure and possible observational signatures of massive WDs \cite{otoniel_effects_2017,otoniel_mass_2021,sousa_prospects_2024}.

The paper is organized as follows. Sec.~\ref{sec:theory} presents the theoretical framework, introducing the two modified dispersion relations and the corresponding WD equation of state used in the stellar calculations. Sec.~\ref{sec:results} discusses the numerical results for the pressure, adiabatic index, and mass radius sequences, with emphasis on the differences between the rational and logarithmic deformation models. Although the stellar sequences are obtained numerically, Appendix~\ref{app:analytic_eos} collects the analytic EoS expressions used as checks on the calculation. Finally, Sec.~\ref{sec:discussion_conclusions} summarizes the physical interpretation, the allowed mass range, and the main limitations that must be addressed before turning the model into a complete astrophysical description.

\section{Theoretical framework}\label{sec:theory}

\subsection{Modified dispersion relations}

The theoretical model has two ingredients. The first one is a deformation of the electron relation between energy and momentum. The second one is the ordinary WD relation between electron number density and ionic mass density. We introduce them separately because the deformation changes the electron pressure, whereas the inertia of the star remains dominated by nuclei. This distinction is essential for the interpretation of all equations of state and stellar sequences shown below.

We start from a deformed relation between energy and momentum written as
\begin{equation}
E^2 f(x)^2-k^2c^2g(x)^2=m^2c^4,
\label{eq:general_mdr}
\end{equation}
where $E$ is the particle energy, $k$ is the magnitude of the particle momentum, $m$ is the particle rest mass, and $c$ is the speed of light. The functions $f$ and $g$ encode the deformation and depend on the dimensionless variable
\begin{equation}
x=\frac{\lambda E}{E_p}.
\end{equation}
Here $E_p$ is the Planck energy and $\lambda$ is a phenomenological deformation parameter. The ordinary relativistic relation must be recovered when the deformation is removed, so the functions satisfy
\begin{equation}
\lim_{x\rightarrow 0}f(x)=1,\qquad
\lim_{x\rightarrow 0}g(x)=1,
\end{equation}
which gives $E^2=k^2c^2+m^2c^4$ in the undeformed limit. In the numerical implementation it is useful to work with the dimensionless parameter $\lambda_0$ defined by
\begin{equation}
\frac{\lambda}{E_p}=\frac{\lambda_0}{mc^2},
\qquad
x=\lambda_0\frac{E}{mc^2}.
\label{eq:lambda0_definition}
\end{equation}
For the WD calculation the relevant mass in Eq.~(\ref{eq:lambda0_definition}) is the electron mass $m_e$. Thus $\lambda_0$ measures the deformation strength relative to the electron rest energy, and the quantity $\lambda_0 E/(m_e c^2)$ controls the size of the correction along a stellar sequence.

The dispersion relation is imposed in a local orthonormal frame comoving with the stellar matter. Using the spacetime signature $(-,+,+,+)$, let $u^\mu$ be the fluid four velocity and $p_\mu$ the electron four momentum. The energy and the squared spatial momentum measured in that frame are
\begin{equation}
E_{\rm loc}=-p_\mu u^\mu,
\qquad
k_{\rm loc}^{\,2}=h^{\mu\nu}p_\mu p_\nu,
\label{eq:local_energy_momentum}
\end{equation}
where
\begin{equation}
h^{\mu\nu}=g^{\mu\nu}+\frac{u^\mu u^\nu}{c^2}
\label{eq:spatial_projector}
\end{equation}
is the projector orthogonal to $u^\mu$. Equation~(\ref{eq:general_mdr}) is therefore understood as a relation between $E_{\rm loc}$ and $k_{\rm loc}$ at each point of the star. The metric $g_{\mu\nu}$ in Eqs.~(\ref{eq:local_energy_momentum}) and (\ref{eq:spatial_projector}) is the ordinary spacetime metric of general relativity and is not identified with the momentum space functions $f$ and $g$.

The first deformation model is
\begin{equation}
f(x)=\frac{1}{1-x},\qquad g(x)=\frac{1}{1-x}.
\label{eq:case1_fg}
\end{equation}

This model was proposed by Magueijo and Smolin in \cite{Magueijo:2001cr} in one of the first scenarios that allows a deformation instead of a violation of Lorentz symmetry with an invariant energy scale. It presents the quantum gravity energy scale as an upper limit attainable by elementary particles. It is also related to the idea that the Planck energy acts as an upper bound on the energies of fundamental particles, complementing the role of the Planck length as a lower bound on the resolution of spacetime distances. This is a common intuition within the quantum gravity community. We are referring to this proposal as rational model.

The rational choice follows from a nonlinear realization of Lorentz transformations on momentum space in which an energy scale accompanies $c$ as an observer independent quantity \cite{Magueijo:2001cr}. In the present calculation, this realization supplies the electron energy $E_e(k)$ and its group velocity. We do not use it to replace the spacetime metric or the gravitational field equations.

Solving Eq.~(\ref{eq:general_mdr}) for the branch with positive energy gives
\begin{equation}
E_{\rm MDR}^{(1)}(k)=A_1+A_3\sqrt{k^2c^2+\bar{m}^{\,2}c^4},
\label{eq:case1_energy}
\end{equation}
where the coefficients are
\begin{equation}
A_1=-\frac{(\lambda/E_p)m^2c^4}
{1-(\lambda^2/E_p^2)m^2c^4},
\end{equation}
\begin{equation}
\bar{m}^{\,2}=
\frac{m^2}
{1-(\lambda^2/E_p^2)m^2c^4},
\end{equation}
and
\begin{equation}
A_3=
\left[1-(\lambda^2/E_p^2)m^2c^4\right]^{-1/2}.
\end{equation}
When $\lambda\rightarrow 0$, one has $A_1\rightarrow 0$, $\bar{m}\rightarrow m$, and $A_3\rightarrow 1$, so Eq.~(\ref{eq:case1_energy}) reduces to the usual relativistic energy. In terms of $\lambda_0$, this branch is real for $|\lambda_0|<1$ for the particle species used in the deformation.

The second model is also motivated by deformed Poincaré symmetries with an invariant energy scale \cite{Amelino-Camelia:1997wnq,Amelino-Camelia:1999iec}. It uses the exponential choice
\begin{equation}
f(x)=\frac{\exp(x)-1}{x},\qquad g(x)=1.
\label{eq:case2_fg}
\end{equation}
The corresponding branch with positive energy is
\begin{equation}
E_{\rm MDR}^{(2)}(k)=
\frac{
\ln\left[
1+(\lambda/E_p)\sqrt{k^2c^2+m^2c^4}
\right]}
{\lambda/E_p}.
\label{eq:case2_energy}
\end{equation}
This expression also recovers the relativistic energy when $\lambda\rightarrow 0$. The logarithm imposes the condition
\begin{equation}
1+\frac{\lambda}{E_p}\sqrt{k^2c^2+m^2c^4}>0.
\label{eq:case2_domain}
\end{equation}
For WDs this condition is important because electrons can become relativistic, so the ratio $E/(m_e c^2)$ can be large even when the deformation parameter is numerically small. We shall refer to this approach as the logarithmic model.

The exponential functions define a second realization of deformed relativistic kinematics motivated by deformed Poincaré symmetries \cite{Lukierski:1991pn,Amelino-Camelia:1997wnq,Amelino-Camelia:1999iec}. This realization is closely related to proposals that introduce an invariant energy scale associated with quantum gravity. It is also connected to the idea of boost saturation in momentum space, according to which the Planck scale constitutes a fundamental upper bound on the energies or momenta that particles can attain in any inertial frame \cite{Bruno:2001mw}. As in the rational case, we treat this choice as a phenomenological model for the electron mass shell. Neither realization is presented here as the unique prediction of a complete quantum gravity theory; rather, both serve as representative phenomenological parametrizations of possible Planck-scale modifications to relativistic kinematics.

\subsection{Degenerate electron gas and WD matter}

The microscopic pressure support in a cold WD comes from a completely degenerate electron gas \cite{Chandrasekhar:1931ih,Glendenning2000}. We keep the density of states in momentum space unchanged, so the deformation affects the energy of each electron state but not the counting of states. With two spin states, the electron number density is
\begin{equation}
n_e=\frac{k_F^3}{3\pi^2\hbar^3},
\label{eq:ne_kf}
\end{equation}
where $k_F$ is the electron Fermi momentum and $\hbar$ is the reduced Planck constant. The pressure generated by an electron relation $E_e(k)$ is
\begin{equation}
P_e(k_F)=
\frac{1}{3\pi^2\hbar^3}
\int_0^{k_F}
\frac{dE_e}{dk}k^3\,dk.
\label{eq:pressure_integral}
\end{equation}
The derivative $dE_e/dk$ is the electron group velocity. Therefore a modified dispersion relation changes the pressure through both the energy branch and its slope. This point is central for the figures of the equation of state, because the pressure at fixed density is controlled by the group velocity inside the Fermi sea.

The mass density is not computed as $m_e n_e$. In a WD, charge neutrality relates the electron density to the ion density. If the nuclei have mass number $A$ and charge $Z$, the mean molecular weight per electron is $\mu_e=A/Z$, and the mass density is approximated by
\begin{equation}
\rho\simeq \mu_e m_u n_e,
\label{eq:rho_ne}
\end{equation}
where $m_u$ is the atomic mass unit. In this work we use $\mu_e=2$, appropriate for a baseline composed of carbon and oxygen in standard WD models \cite{chabrier_equation_1998,potekhin_equation_2000}. Combining Eqs.~(\ref{eq:ne_kf}) and (\ref{eq:rho_ne}) gives
\begin{equation}
k_F=\hbar
\left(
\frac{3\pi^2\rho}{\mu_e m_u}
\right)^{1/3}.
\label{eq:kf_rho}
\end{equation}
This equation connects the density used in stellar structure to the electron Fermi momentum used in the microscopic integrals.

The total energy density entering the relativistic stellar equations is written as the ion rest energy plus the electron contribution,
\begin{equation}
\varepsilon_{\rm tot}\simeq \rho c^2+\varepsilon_e,
\label{eq:epsilon_total}
\end{equation}
where the electron energy density is
\begin{equation}
\varepsilon_e(k_F)=
\frac{8\pi}{(2\pi\hbar)^3}
\int_0^{k_F}E_e(k)k^2\,dk.
\label{eq:electron_energy_density}
\end{equation}
The term $\rho c^2$ dominates the total energy density for ordinary WDs, but retaining $\varepsilon_e$ keeps the relativistic bookkeeping explicit and ensures that all electron branches are treated consistently.

\subsection{Electron equations of state}

We now define the equations of state used in the stellar calculation. The index $i$ labels the electron branch: $i=0$ denotes the undeformed relativistic relation, $i=1$ denotes the rational deformation in Eq.~(\ref{eq:case1_energy}) evaluated with $m=m_e$, and $i=2$ denotes the logarithmic deformation in Eq.~(\ref{eq:case2_energy}) evaluated with $m=m_e$. This construction follows the same free Fermi gas logic used in compact star applications of modified dispersion relations, but adapted here to the electron sector of WDs \cite{Santos:2021avc,Gregg:2008jb,Amelino-Camelia:2009iag}. We write these branches as $E_e^{(i)}(k)$.

For all three cases, the electron contribution is computed from the same two Fermi gas integrals,
\begin{equation}
\varepsilon_e^{(i)}(k_F)=
\frac{8\pi}{(2\pi\hbar)^3}
\int_0^{k_F}
E_e^{(i)}(k)k^2\,dk,
\label{eq:electron_epsilon_i}
\end{equation}
and
\begin{equation}
P_e^{(i)}(k_F)=
\frac{1}{3\pi^2\hbar^3}
\int_0^{k_F}
\frac{dE_e^{(i)}}{dk}k^3\,dk.
\label{eq:electron_pressure_i}
\end{equation}
Thus the pressure used in the stellar calculation is not the undeformed degeneracy pressure unless $i=0$ or $\lambda=0$. For the two deformed branches, both $\varepsilon_e$ and $P_e$ are recomputed with the corresponding electron energy and group velocity, as in relativistic EoS constructions where the microscopic pressure must be evaluated consistently with the chosen particle spectrum \cite{Glendenning2000,potekhin_equation_2013}. The density relation in Eq.~(\ref{eq:rho_ne}) is kept unchanged, so the WD equation of state is the parametric relation
\begin{equation}
\left[
\varepsilon_{\rm tot}^{(i)}(k_F),
P_e^{(i)}(k_F)
\right],
\qquad
\varepsilon_{\rm tot}^{(i)}(k_F)
\simeq
\rho(k_F)c^2+\varepsilon_e^{(i)}(k_F),
\label{eq:parametric_wd_eos}
\end{equation}
with $\rho(k_F)$ fixed by Eqs.~(\ref{eq:ne_kf}) and (\ref{eq:rho_ne}). This form makes explicit what is modified and what is not: the electron energy density and pressure change with the dispersion relation, while the map from electron density to mass density remains the usual WD map.

\subsection{Stellar structure}

At the macroscopic level, the electron degrees of freedom are incorporated into the functions $\varepsilon_{\rm tot}$ and $P$ defined by the equation of state. We assume that the resulting matter is an isotropic perfect fluid with energy momentum tensor
\begin{equation}
T^{\mu\nu}=\frac{\varepsilon_{\rm tot}+P}{c^2}u^\mu u^\nu+P g^{\mu\nu}.
\label{eq:perfect_fluid_tensor}
\end{equation}
The gravitational sector is defined by the Einstein equations
\begin{equation}
G^{\mu\nu}=\frac{8\pi G}{c^4}T^{\mu\nu},
\label{eq:einstein_equations}
\end{equation}
together with $\nabla_\mu T^{\mu\nu}=0$. Under static spherical symmetry, these equations lead to the standard TOV system. The modified electron spectrum enters through $\varepsilon_{\rm tot}(P)$ and does not add a term to the geometric side of Eq.~(\ref{eq:einstein_equations}). This is the same separation between microphysical input and gravitational equilibrium used in the preceding free Fermi gas application \cite{Santos:2021avc}.

The stellar interior is described by the usual static and spherically symmetric metric of general relativity. At the surface, it is matched to the exterior Schwarzschild solution. The functions $f$ and $g$ of the electron dispersion relation are not inserted into either metric.

The final theoretical step is to connect the equation of state to hydrostatic equilibrium. We use the standard TOV equations and introduce no modification of Einstein gravity \cite{Tolman1939,Oppenheimer:1939ne}. The enclosed mass $M(r)$ obeys
\begin{equation}
\frac{dM}{dr}=4\pi r^2\rho(r),
\label{eq:tov_mass}
\end{equation}
where $r$ is the radial coordinate and $\rho(r)$ is the mass density. The pressure profile satisfies
\begin{widetext}
\begin{equation}
\frac{dP}{dr}
=
-\frac{G M(r)\rho(r)}{r^2}
\left(1+\frac{P(r)}{\rho(r)c^2}\right)
\left(1+\frac{4\pi r^3P(r)}{M(r)c^2}\right)
\left(1-\frac{2GM(r)}{rc^2}\right)^{-1}.
\label{eq:tov_pressure}
\end{equation}
\end{widetext}
with $\rho=\varepsilon/c^2$. The three correction factors in Eq.~(\ref{eq:tov_pressure}) are the usual relativistic contributions from pressure inertia, pressure as a source of gravity, and compactness. Since WDs are weakly compact compared with neutron stars, these corrections are small, but using the TOV system keeps the treatment consistent with relativistic WD calculations and with the reference compact star calculation \cite{rotondo_relativistic_2011,boshkayev_general_2013,Santos:2021avc}. The stellar surface is defined by the radius at which the pressure reaches the lower boundary of the tabulated equation of state.

\section{Results}\label{sec:results}

The results are organized to show how the microscopic deformation propagates from the electron gas to the global WD sequence. Instead of emphasizing the absolute EoS, we use the fractional pressure shift as the first diagnostic, because it isolates the part of the pressure that is produced only by the modified electron dispersion relation. We then use the same equations of state in the standard TOV system to obtain the corresponding stellar sequences and maximum masses. Throughout this section, the undeformed case $\lambda_0=0$ is used as the reference curve.

\begin{figure*}
\begin{subfigure}{0.49\textwidth}
\includegraphics[width=\textwidth]{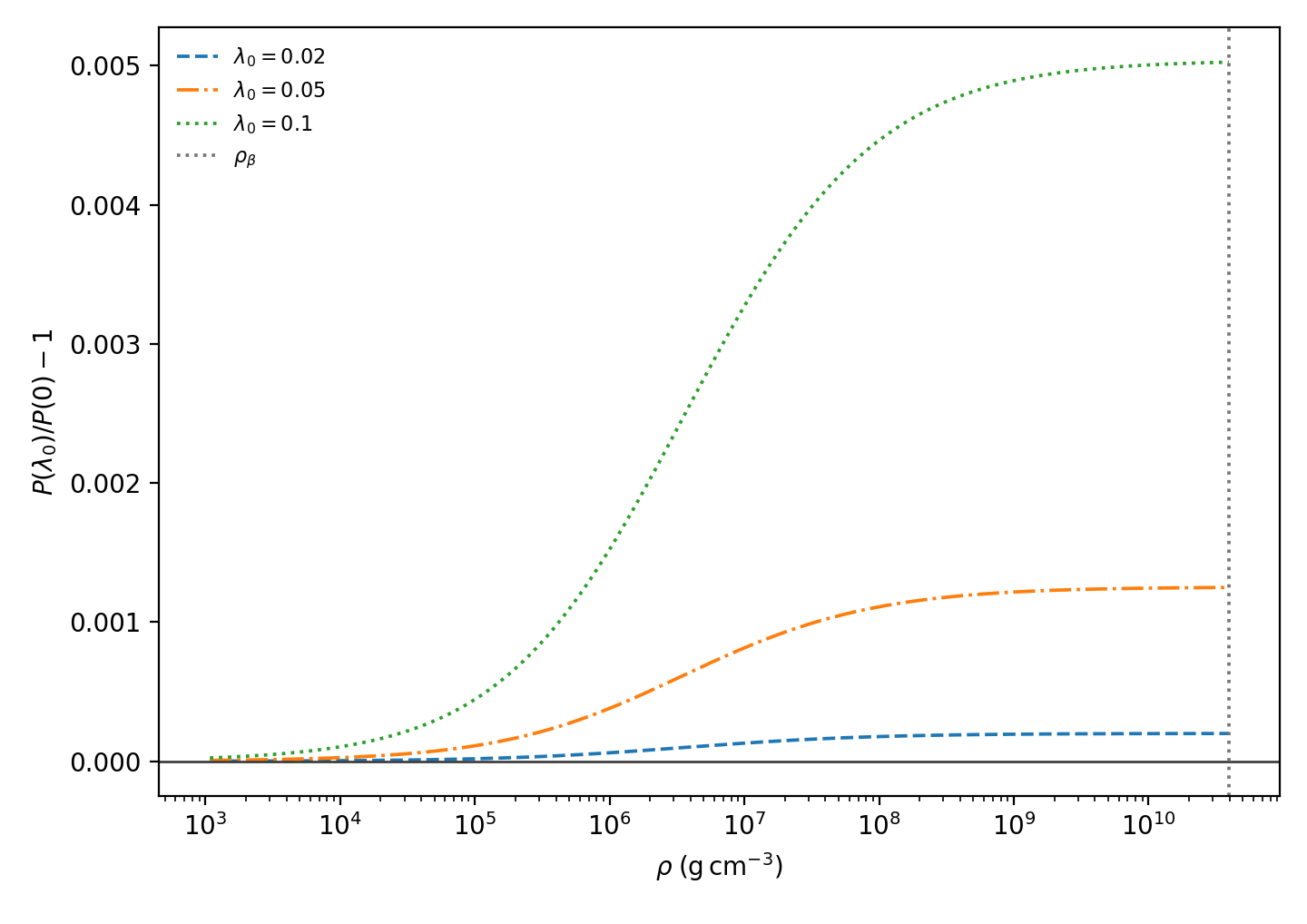}
\caption{Rational model.}
\end{subfigure}
\begin{subfigure}{0.49\textwidth}
\includegraphics[width=\textwidth]{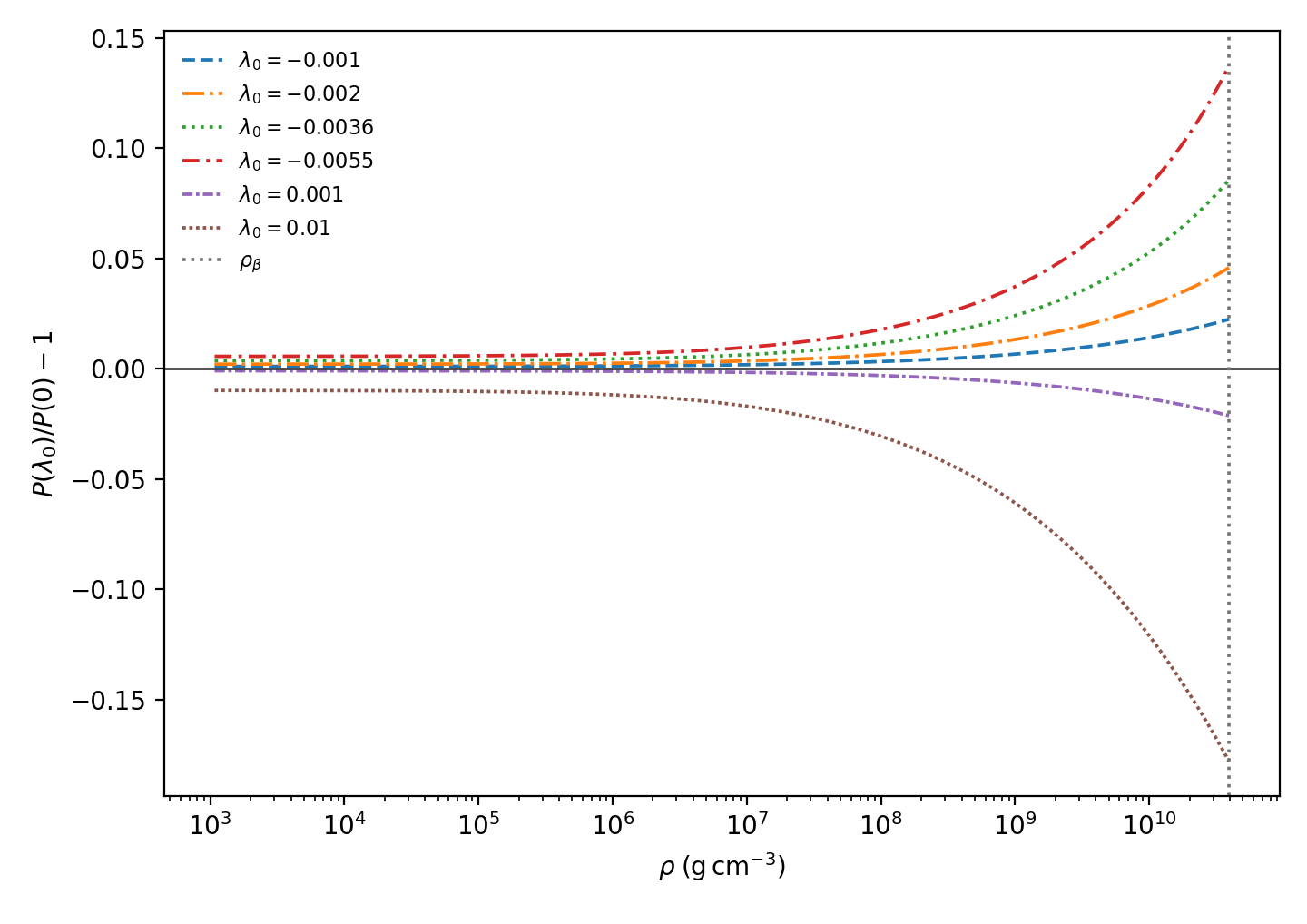}
\caption{Logarithmic model.}
\end{subfigure}
\caption{\label{fig:wd_pressure_fractional}Fractional pressure difference relative to the undeformed EoS. This plot isolates the effect of the modified electron dispersion relation at fixed mass density.}
\end{figure*}

The fractional pressure shift in Fig.~\ref{fig:wd_pressure_fractional} makes the microphysical content clear because the common undeformed background is divided out. The quantity plotted is $P(\lambda_0)/P(0)-1$, so positive values correspond to a stiffer electron gas and negative values correspond to a softer one. This representation shows that the rational model is a controlled and weak deformation over the range considered. One of the reasons for this behavior lies in the fact that although the dispersion relation presents departures from special relativity at first order in the deformation scale \eqref{eq:case1_energy}, the group velocity is only modified at second order. This is responsible for suppressing the pressure corrections in the first scenario.

By contrast, in Fig.~\ref{fig:wd_pressure_fractional}, the logarithmic branch separates the curves much more efficiently because the deformation changes the group velocity through the logarithmic denominator. This is one of the useful outcomes of the model: WDs allow us to see how a change in electron kinematics alone can produce a measurable pressure response without invoking nuclear interactions. We also observe that the central density must remain below the adopted inverse beta decay threshold, $\rho_\beta=3.94\times10^{10}\,{\rm g\,cm^{-3}}$, since above this limit the matter composition is no longer stable \cite{chamel_electron_2015,otoniel_strongly_2019}. The reason for this behavior lies in the fact that the group velocity presents first order corrections in the deformation scale. Besides that, if $\lambda_0$ is positive, the group velocity is reduced in comparison to the special relativistic case, which softens the matter pressure. The opposite behavior happens when $\lambda_0$ is negative, which stiffens the matter contribution.

\begin{figure*}
\begin{subfigure}{0.49\textwidth}
\includegraphics[width=\textwidth]{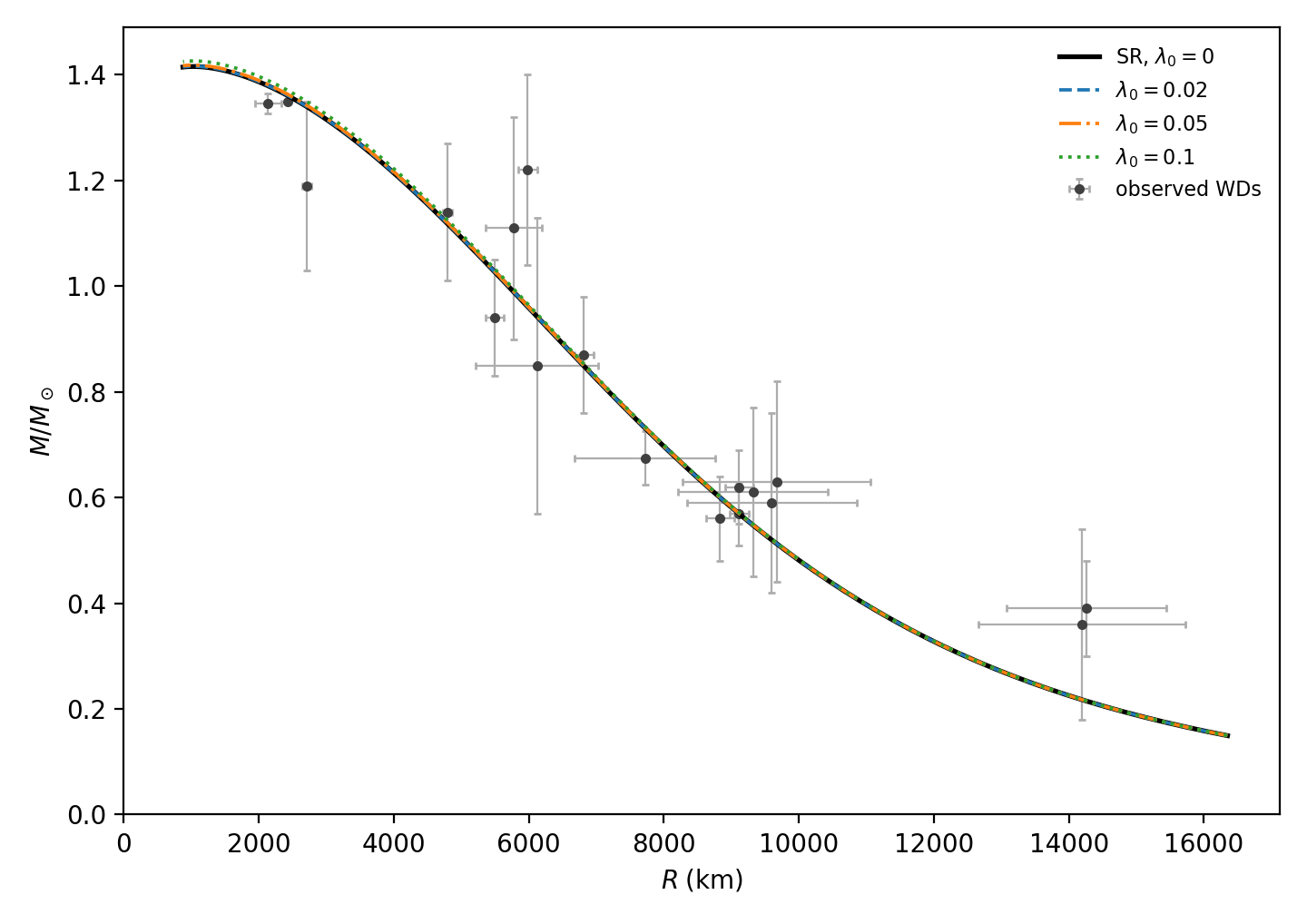}
\caption{Rational model.}
\end{subfigure}
\begin{subfigure}{0.49\textwidth}
\includegraphics[width=\textwidth]{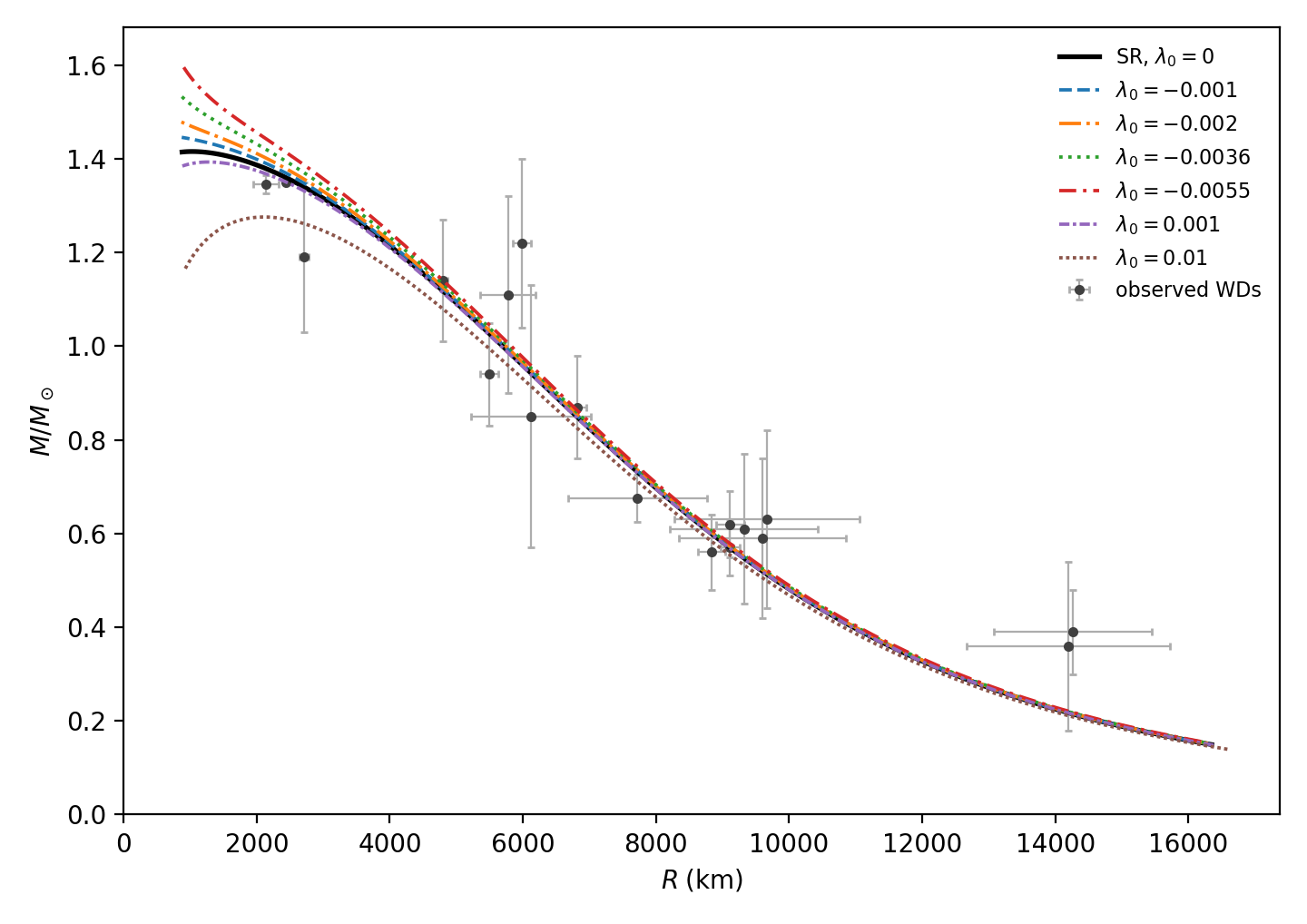}
\caption{Logarithmic model.}
\end{subfigure}
\caption{\label{fig:wd_mr_final}Stellar sequences obtained by coupling the modified electron EoS to the standard TOV equations. The undeformed sequence recovers the Chandrasekhar scale for $\mu_e=2$. Observational white-dwarf masses and radii are shown for comparison, primarily from the catalog of B{\'e}dard, Bergeron, and Fontaine~\cite{Bedard_2017}.}
\end{figure*}

The changes in the EoS propagate into the stellar sequences shown in Fig.~\ref{fig:wd_mr_final}. The rational model produces only a small displacement of the relation between mass and radius, even for the largest positive values considered. This tells us that case 1 is not easily amplified in ordinary WDs when the mass density remains ionic. The logarithmic model behaves differently. Negative values of $\lambda_0$ support larger maximum masses and move the maximum toward smaller radii, which is the expected global response to a stiffer EoS. Positive values soften the sequence and reduce the maximum mass. The model therefore gives a clean sign test: the same WD construction can distinguish whether a deformation increases or decreases the effective electron pressure.

\begin{table}
\caption{\label{tab:summary_final}Summary of the final WD sequences. Masses are in $M_\odot$, radii are evaluated at the maximum mass and given in km, and $\Gamma_{\rm min}$ is the minimum adiabatic index along each tabulated EoS.}
\scriptsize
\begin{ruledtabular}
\begin{tabular}{ccccc}
Case & $\lambda_0$ & $M_{\rm max}$ & $R_{\rm max}$ & $\Gamma_{\rm min}$ \\
\hline
1 & 0.0000 & 1.416 & 1036.6 & 1.334 \\
1 & 0.0200 & 1.416 & 1036.6 & 1.334 \\
1 & 0.0500 & 1.418 & 1037.1 & 1.334 \\
1 & 0.1000 & 1.426 & 1038.6 & 1.334 \\
2 & 0.0000 & 1.416 & 1036.6 & 1.334 \\
2 & -0.0010 & 1.446 & 872.9 & 1.340 \\
2 & -0.0020 & 1.479 & 868.0 & 1.344 \\
2 & -0.0036 & 1.535 & 860.3 & 1.349 \\
2 & -0.0055 & 1.608 & 851.2 & 1.355 \\
2 & 0.0010 & 1.393 & 1276.1 & 1.327 \\
2 & 0.0100 & 1.276 & 2088.0 & 1.275 \\
\end{tabular}
\end{ruledtabular}
\end{table}

\begin{figure*}
\begin{subfigure}{0.49\textwidth}
\includegraphics[width=\textwidth]{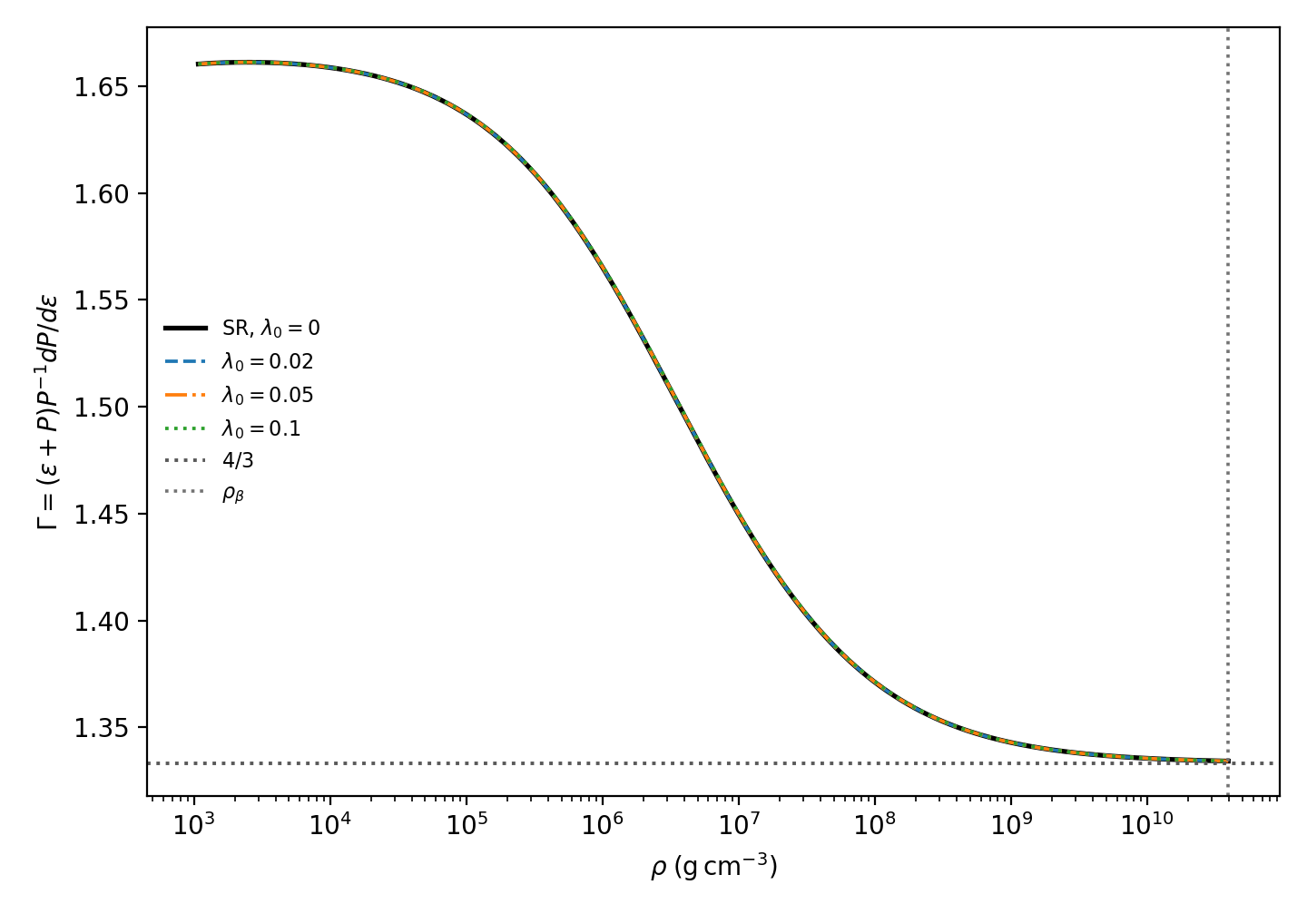}
\caption{Rational model.}
\end{subfigure}
\begin{subfigure}{0.49\textwidth}
\includegraphics[width=\textwidth]{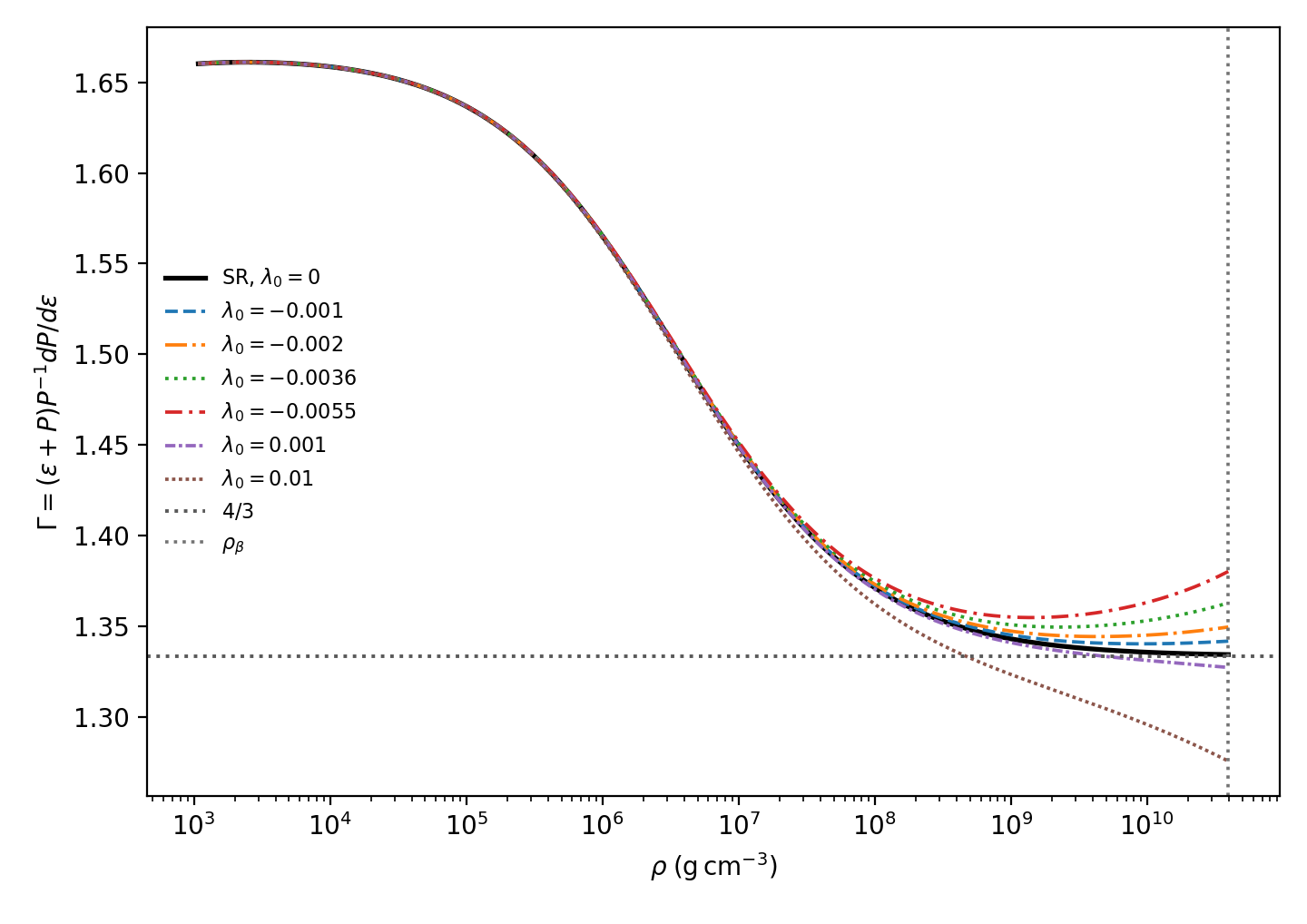}
\caption{Logarithmic model.}
\end{subfigure}
\caption{\label{fig:wd_gamma_final}Adiabatic index computed from the modified EoS. The reference line $\Gamma=4/3$ indicates the usual relativistic stability threshold for a cold degenerate configuration.}
\end{figure*}

The adiabatic index in Fig.~\ref{fig:wd_gamma_final} adds an important physical filter to the mass and radius curves. It is related to the balance between the matter and gravitational pressure. An index $\Gamma\geq 4/3$ is a sign of dynamic stability. The rational model remains close to the undeformed behavior, with the high density limit approaching the expected relativistic value near $4/3$. In the logarithmic model, negative $\lambda_0$ raises the adiabatic index at high density, consistent with the stiffening seen in the EoS. Positive $\lambda_0$ does the opposite and can drive $\Gamma$ below $4/3$, as also summarized in Table~\ref{tab:summary_final}. This does not mean that every negative logarithmic curve is automatically a viable WD model, because the logarithmic domain, the microscopic group velocity and additional WD physics still impose restrictions. It does show, however, that the framework is sensitive to physically meaningful changes in the electron sector and can identify which branches deserve a more detailed observational comparison. For example, this analysis shows that the positive branch of the logarithmic correction is disfavored by this stability criterion of the star in this model. This shows that a detailed analysis has the potential to discard a class of quantum gravity proposals based on such criteria.

\section{\label{sec:discussion_conclusions}Discussion and conclusions}

The main result of this work is that WDs can be used to test modified electron kinematics in a way that is physically cleaner than a direct compact star replacement of one fermion mass by another. In a WD, the pressure is supplied by degenerate electrons, whereas most of the mass density is carried by ions through charge neutrality. The modified dispersion relation therefore changes the pressure integral through the electron energy and group velocity, while the rest mass density remains controlled by the ionic component. This separation is the central point of the model and is also the reason why the response of the star is not just a rescaled version of the neutron gas calculation.

The undeformed sequence gives the normalization of the calculation. For $\mu_e=2$ we obtain $M_{\rm max}=1.416\,M_\odot$ with $R(M_{\rm max})=1036.6$ km, recovering the Chandrasekhar scale within the relativistic WD treatment used here. This agreement is important because all modified sequences in Table~\ref{tab:summary_final} are interpreted relative to this baseline. The rational model, case 1, gives only a small correction over the final range considered. Increasing $\lambda_0$ from 0 to 0.100 raises the maximum mass from $1.416\,M_\odot$ to $1.426\,M_\odot$, while the radius at the maximum changes by only about 2 km. This branch is therefore useful as a controlled deformation, but it does not by itself produce a large observational shift in the WD mass scale.

The logarithmic model, case 2, contains the most relevant phenomenology. Negative values of $\lambda_0$ increase the fractional pressure in Fig.~\ref{fig:wd_pressure_fractional}, stiffen the mass radius sequence in Fig.~\ref{fig:wd_mr_final}, and raise the adiabatic index in Fig.~\ref{fig:wd_gamma_final}. The allowed final sequence of negative values shows a gradual increase of the maximum mass: $1.446\,M_\odot$ for $\lambda_0=-0.001$, $1.479\,M_\odot$ for $\lambda_0=-0.002$, $1.535\,M_\odot$ for $\lambda_0=-0.0036$, and $1.608\,M_\odot$ for $\lambda_0=-0.0055$. These values are not isolated numerical points; they follow the same physical trend. A negative logarithmic deformation increases the electron degeneracy pressure at fixed density, allowing hydrostatic equilibrium to be maintained up to larger masses before the final central density cut is reached.

The physical viability of these larger masses was filtered with two conditions that are directly relevant for cold WDs. First, the adiabatic index must not enter the locally unfavorable regime associated with $\Gamma<4/3$ along the tabulated EoS. Second, the central density of the maximum mass configuration must remain below the adopted inverse beta decay threshold, $\rho_\beta=3.94\times10^{10}\,{\rm g\,cm^{-3}}$. Similar stability filters appear in relativistic studies of rotating and magnetized WDs, where electron captures, pycnonuclear reactions, and turning point criteria restrict the high mass region \cite{boshkayev_maximum_2011,boshkayev_maximum_2013,otoniel_strongly_2019}. Under these two filters, the negative logarithmic branch reaches a possible mass of $1.608\,M_\odot$ at $\lambda_0=-0.0055$, with $\Gamma_{\rm min}=1.355$ and $\rho_c=3.937\times10^{10}\,{\rm g\,cm^{-3}}$. The more conservative point $\lambda_0=-0.0036$ gives $1.535\,M_\odot$, with $\Gamma_{\rm min}=1.349$ and a similar central density below the same threshold. Thus the model naturally produces masses above the Chandrasekhar value, including intermediate values near $1.45\,M_\odot$, $1.48\,M_\odot$, and $1.54\,M_\odot$, before reaching the upper value near $1.6\,M_\odot$ in the final scan.

Positive values of $\lambda_0$ in the logarithmic model give the opposite behavior and help demonstrate that the effect is genuinely tied to the sign of the deformation. For $\lambda_0=0.001$, the maximum mass decreases to $1.393\,M_\odot$ and the minimum adiabatic index falls to $1.327$, already below $4/3$. For $\lambda_0=0.010$, the sequence is much softer, with $M_{\rm max}=1.276\,M_\odot$, $R(M_{\rm max})=2088.0$ km, and $\Gamma_{\rm min}=1.275$. These positive logarithmic cases are therefore not good candidates for stable high mass WDs in the present parameter range. They are still useful because they show that the same formalism can separate stiffening and softening branches without changing the gravitational equations. 

The main conclusion is that a modified electron dispersion relation can produce WD masses above the standard Chandrasekhar scale while keeping the calculation within standard gravity and ordinary charge neutral matter. The strongest final result is the logarithmic negative branch: after imposing the adiabatic stability filter and the inverse beta decay density cut, the model supports possible maximum masses from approximately $1.45\,M_\odot$ up to $1.61\,M_\odot$. This should not be read as a claim that every point in the negative branch is already a complete astrophysical model. Rotation, magnetic fields, Coulomb corrections, crystallization, composition changes, pycnonuclear reactions, and observational selection effects can still modify the final allowed region \cite{otoniel_effects_2017,otoniel_mass_2021,tang_oscillations_2023}. The same mass range can also be approached in other WD scenarios, including modified gravity, magnetic deformation, and compact binary or merger channels \cite{otoniel_white_2026,malheiro_double_2026}. In addition, microscopic constraints on the electron group velocity and on the domain of the logarithmic dispersion relation must be included in a final parameter bound. The result established here is more specific: WDs provide a sensitive laboratory for electron sector MDR effects, and the negative logarithmic branch can raise the maximum mass to the observed super Chandrasekhar range while passing the adiabatic and inverse beta decay filters used in the present calculation.

An alternative extension would promote the momentum dependent functions to an energy dependent spacetime geometry, as in rainbow gravity \cite{Magueijo:2002xx}. That choice would define a different gravitational model and would require a consistent specification of its field equations, conservation law, interior solution, and the energy variable assigned to a many electron system. Since a degenerate star contains occupied states over the full Fermi sea, these ingredients are not fixed by the single particle dispersion relation alone. Deriving stellar equilibrium in such a gravitational extension is left for future work.

\appendix

\section{\label{app:analytic_eos}Analytic electron equation of state expressions}

This appendix lists the analytic electron energy density and pressure used as checks on the numerical EoS. The symbols have the same meaning as in the main text. The stellar sequences are obtained from the defining integrals of Eqs.~(\ref{eq:electron_epsilon_i}) and (\ref{eq:electron_pressure_i}); the formulas below only identify the undeformed gas, case 1, and the perturbative form of case 2.

For the undeformed electron gas, the result is
\begin{widetext}
\begin{equation}
\varepsilon_e^{(0)}(k_F)=
\frac{c}{8\pi^2\hbar^3}
\left[
k_F\sqrt{k_F^2+m_e^2c^2}\left(2k_F^2+m_e^2c^2\right)
-m_e^4c^4
\ln\left(
\frac{k_F+\sqrt{k_F^2+m_e^2c^2}}{m_ec}
\right)
\right],
\end{equation}
\begin{equation}
P_e^{(0)}(k_F)=
\frac{c}{24\pi^2\hbar^3}
\left[
\frac{2k_F^5-m_e^2c^2k_F^3-3m_e^4c^4k_F}
{\sqrt{k_F^2+m_e^2c^2}}
+3m_e^4c^4
\ln\left(
\frac{k_F+\sqrt{k_F^2+m_e^2c^2}}{m_ec}
\right)
\right].
\end{equation}
\end{widetext}

For case 1, define
\begin{equation}
A_{1e}=-\frac{(\lambda/E_p)m_e^2c^4}
{1-(\lambda^2/E_p^2)m_e^2c^4},
\end{equation}
and
\begin{equation}
A_{3e}=\left[1-(\lambda^2/E_p^2)m_e^2c^4\right]^{-1/2},
\end{equation}
and
\begin{equation}
\bar m_e^{\,2}=\frac{m_e^2}
{1-(\lambda^2/E_p^2)m_e^2c^4}.
\end{equation}
The energy density and pressure are
\begin{widetext}
\begin{equation}
\varepsilon_e^{(1)}(k_F)=
A_{1e}
\frac{8\pi}{(2\pi\hbar)^3}\frac{k_F^3}{3}
+A_{3e}\frac{c}{8\pi^2\hbar^3}
\left[
k_F\sqrt{k_F^2+\bar m_e^{\,2}c^2}
\left(2k_F^2+\bar m_e^{\,2}c^2\right)
-\bar m_e^{\,4}c^4
\ln\left(
\frac{k_F+\sqrt{k_F^2+\bar m_e^{\,2}c^2}}{\bar m_ec}
\right)
\right],
\end{equation}
\begin{equation}
P_e^{(1)}(k_F)=
A_{3e}\frac{c}{24\pi^2\hbar^3}
\left[
\frac{2k_F^5-\bar m_e^{\,2}c^2k_F^3-3\bar m_e^{\,4}c^4k_F}
{\sqrt{k_F^2+\bar m_e^{\,2}c^2}}
+3\bar m_e^{\,4}c^4
\ln\left(
\frac{k_F+\sqrt{k_F^2+\bar m_e^{\,2}c^2}}{\bar m_ec}
\right)
\right].
\end{equation}
\end{widetext}

For case 2, write the perturbative form as
\begin{equation}
\varepsilon_e^{(2)}(k_F)=
\varepsilon_e^{(0)}(k_F)+\Delta\varepsilon_e^{(2)}(k_F),
\end{equation}
and
\begin{equation}
P_e^{(2)}(k_F)=
P_e^{(0)}(k_F)+\Delta P_e^{(2)}(k_F),
\end{equation}
where
\begin{widetext}
\begin{align}
\Delta\varepsilon_e^{(2)}(k_F)
=&\frac{c}{8\pi^2\hbar^3}
\left\{
k_F\sqrt{k_F^2+m_e^2c^2}
\frac{\lambda^2c^2}{18E_p^2}
\left[
2k_F^2\left(4k_F^2+7m_e^2c^2\right)+3m_e^4c^4
\right]\right.
\nonumber\\
&\left.
-m_e^6c^6\frac{\lambda^2c^2}{6E_p^2}
\ln\left(
\frac{k_F+\sqrt{k_F^2+m_e^2c^2}}{m_ec}
\right)
-\frac{4\lambda c}{15E_p}k_F^3
\left(3k_F^2+5m_e^2c^2\right)
\right\},
\end{align}
\begin{align}
\Delta P_e^{(2)}(k_F)
=&\frac{c}{24\pi^2\hbar^3}
\left\{
\frac{\lambda^2c^2}{6E_p^2}
\frac{
2k_F^5\left(4k_F^2+5m_e^2c^2\right)
-m_e^4c^4k_F^3
-3m_e^6c^6k_F}
{\sqrt{k_F^2+m_e^2c^2}}
\right.
\nonumber\\
&\left.
+m_e^6c^6\frac{\lambda^2c^2}{2E_p^2}
\ln\left(
\frac{k_F+\sqrt{k_F^2+m_e^2c^2}}{m_ec}
\right)
-\frac{8\lambda c}{5E_p}k_F^5
\right\}.
\end{align}
\end{widetext}
In the numerical sequences, case 2 is evaluated with the exact logarithmic energy of Eq.~(\ref{eq:case2_energy}) inside the defining integrals.

\begin{acknowledgments}
E. Otoniel acknowledges support from FUNCAP under grant BP6-0241-00335.01.00/25. I. P. L. acknowledges partial support from the National Council for Scientific and Technological Development, CNPq, under grant 312547/2023-4. I. P. L. acknowledges the networking support by the COST Action BridgeQG (CA23130), the COST Action RQI (CA23115) and the COST Action FuSe (CA24101) supported by COST (European Cooperation in Science and Technology).
\end{acknowledgments}

\bibliographystyle{apsrev4-2}
\bibliography{wdmdr}

\end{document}